\documentclass{article}
\usepackage{spconf,amsmath,graphicx,amsfonts,epsfig,longtable}
\usepackage{times}
\usepackage{amsbsy}
\usepackage{latexsym}
\usepackage{amssymb}
\title{Study of Buffer-aided physical-layer network coding with ML and MMSE linear code designs for cooperative DS-CDMA
systems \vspace{-0.75em}}

\name{Jiaqi Gu $\maltese$, Rodrigo C. de Lamare $\clubsuit$,
$\maltese$ and Mario Huemer  $\spadesuit$ \vspace{-0.5em}
\thanks{This work is funded by the ESII consortium in the UK under
Task 26 for low-cost wireless ad hoc and sensor
networks.}\vspace{-0.5em}}
\address{\small $\maltese$ ~Communications Research Group, Department of Electronics,
University of York, United Kingdom \\ $\clubsuit$ ~ \small CETUC, PUC-Rio, Rio de Janeiro, Brazil\\
\small $\spadesuit$ Johannes Kepler University Linz, Austria \\
\small Emails: jg849@york.ac.uk, rodrigo.delamare@york.ac.uk,
Mario.Huemer@jku.at \vspace{-1.25em}}

\begin{document}
\ninept \linespread{0.95} \setlength{\abovedisplayskip}{0.4mm}
\setlength{\belowdisplayskip}{0.4mm} \setlength\floatsep{8pt}
\setlength\textfloatsep{8pt} \setlength\intextsep{7.5pt}
\setlength{\abovecaptionskip}{2pt}
\setlength{\belowcaptionskip}{2pt}

\maketitle
\begin{abstract}
In this work, we propose buffer-aided physical-layer network coding
(PLNC) techniques for cooperative direct-sequence code-division
multiple access systems. In the proposed buffer-aided PLNC schemes,
a relay pair selection algorithm is employed to obtain the relay
pair and the packets in the buffer entries with the best performance
and the associated link combinations used for data transmission. We
also devise PLNC schemes based on the design of linear network
coding matrices using maximum likelihood and minimum mean-square
error criteria to generate the network coded symbols that are sent
to the destination. Simulation results show that the proposed
techniques significantly outperform prior art.

\end{abstract}
\vspace{-0.5em}
\section{Introduction}

Interference mitigation is one of the key problems in wireless
communications and many approaches have been devised for this task
in the last decades or so. Unlike traditional strategies that treat
interference as a nuisance to be avoided, physical-layer network
coding (PLNC) techniques take advantage of the superposition of
radio signals and embrace the interference to improve throughput
performance \cite{Zhang}. PLNC techniques have generated a number of
fertile theoretical and application-oriented studies, and are
foreseen to be successfully implemented in future wireless
applications
\cite{Sanna,he7006802,yang7466815,nokleby7182361,zhang7342978}. In
wireless and cognitive radio networks, the physical superposition of
the signals can be seen as a benefit rather than an interference and
exploited for coding at the physical-layer level \cite{Chou}. The
communication can be protected against attacks from malicious nodes
\cite{Jaggi}, eavesdropping entities \cite{Cai}, and impairments
such as noise and information losses \cite{Yeung} thanks to the
property of the network acting as an encoder. In peer-to-peer
networks, the well-known problem of the missing block at the end of
the download can be avoided by the distribution of a number of
encoded versions of the source data \cite{Jain}.

PLNC has significant advantages in wireless multi-hop networks.
Multiple relay nodes are employed in the network to transmit data
from sources to the destination \cite{Burr}. It allows a node to
exploit as far as possible all signals that are received
simultaneously, rather than treating them as interference
\cite{Burr}. Additionally, instead of decoding each incoming data
stream separately, a node detects and forwards the function of the
incoming data streams \cite{Burr1}. There are several different
network coding techniques, namely, the XOR mapping schemes and
linear network coding designs
\cite{Zhang,Ahlswede,Sanna,Li,he7006802,zhang7342978}.

In cooperative relaying systems, strategies that employ relays have
been investigated in
\cite{Jing,Clarke,Talwar,de2012joint,Jiaqi1,Peng11,Jiaqi2}. In order
to further increase the quality and reliability of cooperative
schemes, the concept of buffer has been introduced and used to equip
relay nodes in cooperative relaying scenarios
\cite{Zlatanov1,Zlatanov2,Krikidis,Ikhlef,bfdstc,gu2017buffer}. In
this context, relay selection algorithms must be employed to obtain
the combination of links in the buffer entries that optimize some
criterion and result in performance improvements.

In this work, we propose buffer-aided PLNC techniques for
cooperative direct-sequence code-division multiple access (DS-CDMA)
systems. In the proposed buffer-aided PLNC schemes, a relay pair
selection algorithm is employed to obtain the relay pair and the
packets in the buffer entries with the highest SINR and the
associated link combinations used for data transmission. We also
devise PLNC schemes based on the design of linear network coding
matrices using maximum likelihood (ML) and minimum mean-square error
(MMSE) criteria to generate the network coded symbol (NCS) that is
then sent to the destination. Simulation results show that the
proposed techniques significantly outperform prior art.

This paper is structured as follows: Section 1 introduces the
cooperative DS-CDMA system model. Section 2 presents the proposed ML
and MMSE linear network code designs, whereas Section 3 describes
the proposed buffer-aided PLNC transmission scheme. Section 4
presents the simulations and Section 5 gives the conclusions.

%
%
%

\section{Cooperative DS-CDMA System Model}

\vspace{-0.5em}
\begin{figure}[!htb]
\begin{center}
\def\epsfsize#1#2{0.75\columnwidth}
\epsfbox{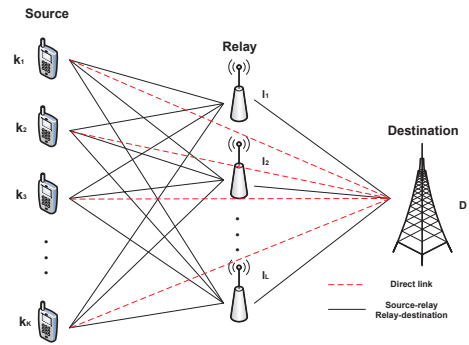} \vspace{-0.75em}\caption{
Uplink of a cooperative DS-CDMA system.}  \label{ch5_fig1}
\end{center}
\end{figure}

We consider the uplink of a two-phase cooperative synchronous
DS-CDMA system with $K$ users, $L$ relays equipped with finite-size
buffers capable of storing $J$ packets and $N$ chips per symbol that
experiences channels with flat fading, as shown in Fig.
\ref{ch5_fig1}. The system is equipped with a cooperative protocol
at each relay and the transmit data are organized in packets of $P$
symbols. The received signals are filtered by a matched filter and
sampled at chip rate to obtain sufficient statistics and organized
into $N \times 1$ vectors $\textbf{y}_{sr}$, $\textbf{y}_{sd}$ and
$\textbf{y}_{rd}$, which represent the signals received from the
sources to the relays, the sources to the destination and the relay
to the destination, respectively. Every set of $m$ users and $m$
relays are randomly assigned into a group, forming pairs of users
and relays. The signal received from the remaining users to the
target relays are seen as interference.

In the first phase, the signals received at the destination and the
$l$-th relay can be described as
\begin{equation}\label{ch5_equation0}
\textbf{y}_{sd}=\sum\limits_{k=1}^K a_{sd}^k\textbf{s}_k
h_{sd,k}b_k+\textbf{n}_{sd},
\end{equation}
\begin{equation}\label{ch5_equation1}
\textbf{y}_{sr_l}=\sum\limits_{k=1}^K a_{sr_l}^k\textbf{s}_k
h_{sr_l,k}b_k+\textbf{n}_{sr_l},
\end{equation}
where $b_k$ correspond to the transmitted symbols. The quantities
$a_{sd}^k$ and $a_{sr_l}^k$ are the $k$-th user's amplitude from the
source to the destination and from the source to relay $l$,
respectively. The vector $\textbf{s}_k=[s_k(1),s_k(2),...s_k(N)]^T$
is the signature sequence for user $k$. The quantities $h_{sd,k}$
and $h_{sr_l,k}$ are the complex channel coefficients from user $k$
to the destination and from user $k$ to the $l$-th relay,
respectively. The vectors $\textbf{n}_{sd}$ and $\textbf{n}_{sr_l}$
are the noise vectors that contain samples of zero mean complex
Gaussian noise with variance $\sigma^2$.

After the detected symbols of interest are generated, a PLNC scheme
is then employed at each relay. When the bit-wise XOR operation (the
modulo-2 sum in the binary field) is considered, we assume that the
mapping from user symbol to a modulated symbol is denoted as
$\mathcal{M}_B$. This mathematical mapping is then expressed as
\begin{equation}\label{ch5_equation2}
b_k= \mathcal{M}_B = 1-2c_k.
\end{equation}
We then perform the above mapping from $\hat{b}_{r_ld,k}$, which
represents the decoded symbol for user $k$ at the output of relay
$l$ using a relay protocol, to its corresponding user symbol as
described by
\begin{equation}\label{ch5_equation3}
\hat{b}_{r_ld,k} \rightarrow c_l^k.
\end{equation}
Similarly, when linear network coding is adopted at the $l$-th
relay, the corresponding mapping $\mathcal{L}_B$ is given by
\begin{equation}\label{ch5_equation4}
b_l=\mathcal{L}_B=\sum\limits_{k=1}^{K}g_{kl}\hat{b}_{r_ld,k}.
\end{equation}
In the following, we briefly review existing PLNC schemes that
employ 
linear network codes at each relay.
When the signals are sent to the relays, we obtain the detected
symbol as given by
\begin{equation}\label{ch5_equation5}
\hat{b}_{r_{\Omega(l)}d,\Upsilon(k)}=Q \Big( (
\textbf{w}_{s_{\Upsilon(k)}r_{\Omega(l)}} )^H \textbf{y}_{sr_l}
\Big),
\end{equation}
where $\Upsilon$ is a $1 \times m$ user set and $\Omega$ is the
corresponding $1 \times m$ relay set since we combine every $m$
users and $m$ relays into a sub transmission group.
$\hat{b}_{r_{\Omega(l)}d,\Upsilon(k)}$ is the detected result for
user $\Upsilon(k)$ at relay $\Omega(l)$,
$\textbf{w}_{s_{\Upsilon(k)}r_{\Omega(l)}}$ refers to the detector
for user $\Upsilon(k)$ at relay $\Omega(l)$. After that, the
detected symbol is mapped through (\ref{ch5_equation2}) to obtain
\begin{equation}\label{ch5_equation6}
c_{\Omega(l)}^{\Upsilon(k)}=(1-\hat{b}_{r_{\Omega(l)}d,\Upsilon(k)})/2.
\end{equation}
If linear network coding techniques
\cite{cai2011switched,uchoa2011design,healy2016design} are adopted
at the relays, every relay group will be allocated a unique $m
\times m$ linear network coding matrix $\textbf{G}$ whose structure
is given by
\begin{equation}\label{ch5_equaion16}
\textbf{G} = \left[\begin{array}{c c c c}
g_{11} & g_{12} \ \ \dots \ \ g_{1m} \\
g_{21} & g_{22} \ \ \dots \ \ g_{2m}\\
&\vdots\\
g_{m1} & g_{m2} \ \ \dots \ \ g_{mm}\\
 \end{array}\right].
\end{equation}
Thus, the following linear combination can be obtained as
\begin{equation}\label{ch5_equation17}
b_{\Omega(l)}= \sum\limits_{k=1}^{m}   g_{kl}\
\hat{b}_{r_{\Omega(l)}d,\Upsilon(k)},
\end{equation}
where $\hat{b}_{r_{\Omega(l)}d,\Upsilon(k)}$ is the detected symbol
for user $\Upsilon(k)$ at relay $\Omega(l)$. After
that, in order to forward the encoded information to the
destination, the following mapping operation is required
\begin{equation}\label{ch5_equation8}
b_{\Omega(l)}=1-2c_{\Omega(l)}.
\end{equation}
Consequently, in the second phase the encoded symbols are stored and
prepared to be sent to the destination. The NCS are then transmitted
from relay set $\Omega$ to the destination as described by
\begin{equation}\label{ch5_equation9}
\textbf{y}_{r_{\Omega}d}=\sum\limits_{l=1}^{m}
\textbf{h}_{r_{\Omega(l)}d} b_{\Omega(l)}+\textbf{n}_{rd},
\end{equation}
where $\textbf{h}_{r_{\Omega(l)}d}=a_{r_{\Omega(l)}d}^{\Upsilon}
\textbf{s}_{\Upsilon} h_{r_{\Omega(l)}d}$ denotes the $N \times 1$
channel vector from relay $\Omega(l)$ to the destination,
$a_{r_{\Omega(l)}d}^{\Upsilon}$ is the amplitude for combined source
(user set $\Upsilon$) from the $\Omega(l)$-th relay to the
destination, $h_{r_{\Omega(l)}d}$ is the complex channel fading
coefficient from the $\Omega(l)$-th relay to the destination,
$\textbf{s}_{\Upsilon}$ is the $N \times 1$ spreading sequence
assigned to the NCS $b_{\Omega(l)}$ and $\textbf{n}_{rd}$ is the
$N\times1$ zero mean complex Gaussian noise with variance
$\sigma^2$.


At the destination, the detected symbol for users from set
$\Upsilon$ after linear network coding at each of the relays is
obtained by
\begin{equation}\label{ch5_equation19}
\hat{b}_{\Omega(l)}=Q \Big( \textbf{G}^{-1}
(\textbf{w}_{r_{\Omega}d})^H \textbf{y}_{r_{\Omega}d} \Big).
\end{equation}
The decoding process for linear network coding is expressed by
\begin{equation}\label{ch5_equation20}
\begin{array}{rl}
\hat{b}_{\Omega(1)}&=g_{\Upsilon(1) \Omega(1)}\hat{b}_{\Upsilon(1)} + g_{\Upsilon(2) \Omega(1)} \hat{b}_{\Upsilon(2)}+...+ g_{\Upsilon(m) \Omega(1)} \hat{b}_{\Upsilon(m)}\\
&\ \  \ \ \ \ \ \ \ \ \ \ \ \ \ \ \ \ \ \ \ \ \ \ \ \ \ \ \ \ \ \ \vdots  \  \\
\hat{b}_{\Omega(m)}&=g_{\Upsilon(1) \Omega(m)}\hat{b}_{\Upsilon(1)} + g_{\Upsilon(2) \Omega(m)} \hat{b}_{\Upsilon(2)}+...+ g_{\Upsilon(m) \Omega(m)} \hat{b}_{\Upsilon(m)}\\
\end{array}
\end{equation}
The $m$ unknown detected symbols $\hat{b}_{\Upsilon(k)}, k=1,2,...m$
for each user in user set $\Upsilon$ can be obtained by solving the
above $m$ equations. However, in this scenario, since we have direct
transmission, there is another decoding alternative. Specifically,
we can obtain the detected symbol for the user
$\hat{b}_{\Upsilon(k)}$ as given by:
\begin{equation}\label{ch5_equation21}
\hat{b}_{\Upsilon(k)}=\frac{\hat{b}_{\Omega(l)}-\big(
g_{\Upsilon(1)\Omega(l)}\hat{b}^{\Upsilon(1)}_{sd} + \ldots +
g_{\Upsilon(m)\Omega(l)} \hat{b}^{\Upsilon(m)}_{sd}
\big)}{g_{\Upsilon(k)\Omega(l)}}.
\end{equation}
With direct transmissions, we can effectively reduce the bit error
rate. In particular, there is a possibility that the detection for
$\hat{b}_{\Omega(l)}, l \in \Omega$ is incorrect, which could
directly lead to a problem when solving the system of equations in
(\ref{ch5_equation20}), causing incorrect decisions. However, this
is not a problem when applying (\ref{ch5_equation21}) as we can
select another detected symbol $\hat{b}_{\Omega(s)}, s\neq l$ as the
minuend in (\ref{ch5_equation21}) instead of the possibly incorrect
detected symbol $\hat{b}_{\Omega(l)}$.

\section{Proposed Cooperative Linear Network Coding Schemes}

In this section, we present novel linear network coding designs for
$\textbf{G}$. The idea is inspired by the work in \cite{Huemer} but
applied in a completely different context. In the cooperative PLNC
transmission scenario, the simplest way to construct $\textbf{G}$ is
to generate it randomly \cite{Ho1}, however, this approach often
does not result in the optimum performance as it does not satisfy an
optimality criterion. Therefore, in order to obtain performance
advantages, we devise methods based on the ML and MMSE criteria.

\subsection{Maximum Likelihood Design}

When a linear network coding scheme is adopted at the relay, we
introduce the $m \times m$ linear network code matrix $\textbf{G}$,
which is used to generate the NCS. An ML design of $\textbf{G}$ that
requires the evaluation of all possible ($2^{m^2}$) binary matrices
is given by
\begin{equation}\label{ch5_equation22}
\begin{split}
\hspace{-0.5em} \textbf{G}^{\rm ML}={\rm arg}& \min _{1\leq j\leq
2^{m^2}} \Big\|
  \left[\hspace{-0.3em}
 \begin{array}{l}
   b_{\Omega(1)}\\
   \ \ \ \vdots \\
   b_{\Omega(m)}\\
 \end{array} \hspace{-0.3em} \right] - \textbf{G}^{-1}
 \left[\hspace{-0.3em}
 \begin{array}{l}
  \textbf{w}_{r_{\Omega(1)}d}^H \textbf{y}_{r_{\Omega(1)}d}\\
  \ \ \ \ \ \ \ \ \ \ \ \vdots\\
  \textbf{w}_{r_{\Omega(m)}d}^H \textbf{y}_{r_{\Omega(m)}d} \\
 \end{array}\hspace{-0.5em} \right] \Big\|^2\\,
 \end{split}
\end{equation}
where $\textbf{w}_{r_ld}$ is the linear detector used at relay $l$,
$\textbf{y}_{r_ld}$ is the signal transmitted from relay $l$. In
summary, the principle of this approach is to obtain an $m \times m$
code matrix $\textbf{G}^{\rm ML}$ to match $\textbf{G}$ (or
$\textbf{G}^{\rm ML})$. Thus, the minimum distance between the
transmitted symbols and detected symbols can be obtained. Note that
this approach corresponds to a combinatorial problem where we must
test all $2^{m^2}$ possible candidates to obtain $\textbf{G}^{\rm
ML}$ associated with the best performance. The computational cost is
one of the disadvantages of this approach but the design of
$\textbf{G}^{\rm ML}$ can be carried out online. 
At the destination side, we employ the same matrix $\textbf{G}^{\rm
ML}$ that is applied at the relays in order to obtain better
detection performance.

\subsection{MMSE design}

The ML approach in (\ref{ch5_equation22}) does not consider the
influence brought by noise. This fact motivates us to seek another
efficient decoding method that can consider both interference and
noise. In order to further exploit the minimum distance and improve
the transmission performance, we introduce a code matrix
$\tilde{\textbf{G}}$ at the destination to match the generated
binary matrix $\textbf{G}$ and perform the linear network coding
operation. The proposed MMSE linear network coding matrix
$\textbf{G}^{\rm MMSE}$ is obtained as follows:
\begin{equation}\label{ch5_equation23}
 \begin{split}
\hspace{-0.5em} \textbf{G}^{\rm MMSE}={\rm arg}& \min
_{\tilde{\textbf{G}}} \ E \left[ \Big\|
  \left[\hspace{-0.3em}
 \begin{array}{l}
   b_{\Omega(1)}\\
   \ \ \ \vdots\\
   b_{\Omega(m)}\\
 \end{array} \hspace{-0.3em} \right] - \tilde{\textbf{G}}
 \left[\hspace{-0.3em}
 \begin{array}{l}
  \textbf{w}_{r_{\Omega(1)}d}^H \textbf{y}_{r_{\Omega(1)}d}\\
  \ \ \ \ \ \ \ \ \ \ \ \vdots\\
  \textbf{w}_{r_{\Omega(m)}d}^H \textbf{y}_{r_{\Omega(m)}d} \\
 \end{array}\hspace{-0.5em} \right]\Big\|^2\right],
 \end{split}
\end{equation}
where ideally we have
$b_{\Omega(1)}^{\Upsilon(k)}=b_{\Omega(2)}^{\Upsilon(k)}=...=b_{\Omega(m)}^{\Upsilon(k)},
k \in [1,m]$. This design problem can be recast as the following:
\begin{equation}\label{ch5_equation24}
\textbf{G}^{\rm MMSE}={\rm arg} \ \min _{\tilde{\textbf{G}}}
E\Big[\Big\| \textbf{a}-\tilde{\textbf{G}}\textbf{b} \Big\|^2\Big],
\end{equation}
where the quantities in the argument are $\textbf{a} =
[b_{\Omega(1)}~\ldots~b_{\Omega(m)}]^T$ and $\textbf{b} =
[\textbf{w}_{r_{\Omega(1)}d}^H \textbf{y}_{r_{\Omega(1)}d} ~\ldots~
\textbf{w}_{r_{\Omega(m)}d}^H \textbf{y}_{r_{\Omega(m)}d}]^T$.
By taking the gradient of the cost function with respect to
$\tilde{\textbf{G}}$ and equating the terms to zero, we obtain the
MMSE linear network code matrix given by
\begin{equation}\label{ch5_equation27}
\begin{split}
{\mathbf G}^{\rm MMSE}& ={\mathbf P}_{ab}
{\mathbf R}_b^{-1},
\end{split}
\end{equation}
where the statistical quantities in the MMSE code matrix are the
cross-correlation matrix $\mathbf{P}_{ab}=
E[\mathbf{a}\mathbf{b}^H]$ and the auto-correlation matrix ${\mathbf
R}_b= E[\mathbf{b}\mathbf{b}^H]$. The elements of $\mathbf{P}_{ab}$
are given by $[\mathbf{P}_{ab}]_{k,j} = (\sum_{i=1}^m g_{ik}g_{ik}
\sigma_{i}^2){\mathbf h}^H_{{\mathbf r}_{\Omega(j)d}} {\mathbf
w}_{{\mathbf r}_{\Omega(j)d}}$, for $k,j = 1, \ldots, m$, whereas
the main diagonal entries of ${\mathbf R}_b$ are described by
$[{\mathbf R}_b]_{j,j} = {\mathbf w}^H_{{\mathbf r}_{\Omega(j)d}}
({\mathbf h}_{{\mathbf r}_{\Omega(j)d}}{\mathbf h}^H_{{\mathbf
r}_{\Omega(j)d}}+ \sigma^2{\mathbf I}) {\mathbf w}_{{\mathbf
r}_{\Omega(j)d}}$, for $j = 1, \ldots, m$. \vspace{-1em}

\section{Proposed Buffer-Aided PLNC Scheme}

In this section, we consider groups of $m=2$ users and $m=2$ relays
and develop a buffer-aided PLNC scheme, where each relay is equipped
with a buffer so the received data can be stored and wait until the
link pair associated with the best performance is selected.
Consequently, encoded data are stored at the buffer entries and then
forwarded to the destination when the appropriate time interval
comes. Moreover, the destination is also equipped with a buffer so
that the detected symbols from the direct transmissions can be
stored. After that, the detected symbols are obtained through PLNC
decoding and mapping operations at the appropriate time instants.
\\
\\
\textbf{Transmission between a selected user-relay pair ($m=2$)}
\\ \\
At the relays, the received signal is processed by linear network
coding and forwarded to the destination, which employs a decoding
matrix and a linear receiver. Specifically, the proposed algorithm
starts with a selection procedure using all possible link
combinations associated with all relay pairs from both source-relay
and relay-destination phases. Since any linear network coding
technique can be adopted, every ($m=2$) relays are combined into a
group and paired with ($m=2$) users. Note that arbitrary numbers of
users and relays can also be considered but for the sake of
simplicity, we only consider $m=2$ relays and users. In this case,
the signal transmitted from other users are seen as the interference
component. All possible links are considered and their SINR are then
calculated:{
\begin{equation}\label{ch5_equation31}
{\rm SINR}_{sr_{\Omega}}=\frac{\sum\limits_{k=1}^K
\sum\limits_{l=1}^{m} \textbf{w}_{s_kr_{\Omega(l)}}^H
\rho_{s_kr_{\Omega(l)}}
\textbf{w}_{s_kr_{\Omega(l)}}}{\sum\limits_{k=1}^K
\sum\limits_{\substack{j=1\\j \not\in \Omega}}^L
\textbf{w}_{s_kr_j}^H \rho_{s_kr_j} \textbf{w}_{s_kr_j} +
\sum\limits_{l=1}^{m} \sigma^2\textbf{w}_{s_kr_{\Omega(l)}}^H
\textbf{w}_{s_kr_{\Omega(l)}}},
\end{equation}
\begin{equation}\label{ch5_equation32}
{\rm SINR}_{r_{\Omega}d}=\frac{\sum\limits_{k=1}^K
\sum\limits_{l=1}^{m} (\textbf{w}_{r_{\Omega(l)}d}^k)^H
\rho_{r_{\Omega(l)}d}^k
\textbf{w}_{r_{\Omega(l)}d}^k}{\sum\limits_{k=1}^K
\sum\limits_{\substack{j=1\\j \not\in \Omega}}^L
(\textbf{w}_{r_jd}^k)^H \rho_{r_jd}^k \textbf{w}_{r_jd}^k +
\sum\limits_{l=1}^{m} \sigma^2(\textbf{w}_{r_{\Omega(l)}d}^k)^H
\textbf{w}_{r_{\Omega(l)}d}^k},
\end{equation}
where $\rho_{s_kr_{\Omega(l)}}=\textbf{h}_{s_kr_{\Omega(l)}}^H
\textbf{h}_{s_kr_{\Omega(l)}}$ is the correlation coefficient of the
desired user $k$ between the source and relay $\Omega(l)$,
$\rho_{r_{\Omega(l)}d}^k=(\textbf{h}_{r_{\Omega(l)}d}^k)^H
\textbf{h}_{r_{\Omega(l)}d}^k$ is the correlation coefficient for
user $k$ from relay $\Omega(l)$ to the destination.
$\textbf{h}_{s_kr_{\Omega(l)}}=a_{s_kr_{\Omega(l)}}\textbf{s}_kh_{s_kr_{\Omega(l)}}$
is the channel vector from user $k$ to relay $\Omega(l)$. In
(\ref{ch5_equation31}), ${\rm SINR}_{sr_{\Omega}}$ denotes the SINR
for the combined paths from all users to relay set $\Omega$,
$\textbf{w}_{s_kr_{\Omega(l)}}$ is the detector used at the relay
$\Omega(l)$. When the RAKE receiver is adopted at the corresponding
relay, $\textbf{w}_{s_kr_{\Omega(l)}}$ is expressed as
\begin{equation}\label{ch5_equation33}
\textbf{w}_{s_kr_{\Omega(l)}}=\textbf{h}_{s_kr_{\Omega(l)}}.
\end{equation}
Similarly, if the linear MMSE receiver
\cite{RCDL5,de2007reduced,de2010reduced,uchoa2016iterative,ruan2016robust}
is employed at the relays, $\textbf{w}_{s_kr_{\Omega(l)}}$ is equal
to
\begin{equation}\label{ch5_equation34}
\textbf{w}_{s_kr_{\Omega(l)}}=\bigg(\sum\limits_{k=1}^K
\textbf{h}_{s_kr_{\Omega(l)}}\textbf{h}^H_{s_kr_{\Omega(l)}}+\sigma^2
\textbf{I}\bigg)^{-1}\textbf{h}_{s_kr_{\Omega(l)}},
\end{equation}
$\textbf{h}_{s_kr_{\Omega(l)}}=a_{s_kr_{\Omega(l)}}\textbf{s}_k
h_{s_kr_{\Omega(l)}}$ is the effective signature vector from user
$k$ to the relay $\Omega(l)$. In (\ref{ch5_equation32}), ${\rm
SINR}_{r_{\Omega}d}$ represents the SINR for the combined paths from
relay set $\Omega$ to the destination. The receive filter
$\textbf{w}_{r_{\Omega(l)}d}^k$ is employed by the detector used at
the destination. When the RAKE receiver is adopted at the
destination, $\textbf{w}_{r_{\Omega(l)}d}^k$ is expressed as
\begin{equation}\label{ch5_equation35}
\textbf{w}_{r_{\Omega(l)}d}^k=\textbf{h}_{r_{\Omega(l)}d}^k.
\end{equation}
In an analog way, if the linear MMSE receiver is employed at the
relays, $\textbf{w}_{r_{\Omega(l)}d}^k$ is equal to
\begin{equation}\label{ch5_equation36}
\textbf{w}_{r_{\Omega(l)}d}^k=\bigg(\sum\limits_{k=1}^K
\textbf{h}_{r_{\Omega(l)}d}^k(\textbf{h}_{r_{\Omega(l)}d}^k)^{H}+\sigma^2
\textbf{I}\bigg)^{-1} \textbf{h}_{r_{\Omega(l)}d}^k.
\end{equation}
Both RAKE and MMSE receivers are considered here due to their
reasonably low complexity and it should be mentioned that other
detectors \cite{RCDL4,itic,li2011multiple,mbdf,cai2015adaptive}
including the ML detector can also be used. Alternatively, transmit
processing techniques can also be employed
\cite{de2013massive,zhang2015large,zu2013generalized,zu2014multi}.

After the computation of receive filters and SINR values, we sort
all these results according to a decreasing power level and choose
the relay pair ($m=2$) with the highest SINR as given by
\begin{equation}\label{ch5_equation37}
{\rm SINR_{i,j}={\rm arg}\max _{\substack{ {\Omega} \in
[1,2,...,L]}} \{ {\rm SINR}_{sr_{\Omega}},{\rm SINR}_{r_{\Omega}d}
\}},
\end{equation}
where ${\rm SINR_{i,j}}$ denotes the highest SINR associated with
the relay $i$ and relay $j$. After the relay pair ($m=2$) with the
highest SINR is selected, the signal is transmitted through the
corresponding channels. In this case, two different situations of
the buffer mode need to be considered as follows.
\\
\\
\textbf{Transmission mode}:\\\\
If the link combinations associated with the selected relay set
belongs to the relay-destination phase, the buffers are turned to
the transmission mode. A buffer space check needs to be conducted
first to ensure the corresponding buffer entries are not empty,
namely:
\begin{equation}\label{ch5_equation38}
\Phi^{\rm buffer}_{i}\neq\varnothing, \  i\in [1,2,...,L],
\end{equation}
and
\begin{equation}\label{ch5_equation39}
\Phi^{\rm buffer}_{j}\neq\varnothing, \ j\in [1,2,...,L],
\end{equation}
where $\Phi^{\rm buffer}_{i}$ and $\Phi^{\rm buffer}_{j}$ denote the
buffers equipped at relay $i$ and relay $j$. If the corresponding
buffer entries are not empty, we transmit the NCS according to
(\ref{ch5_equation9}). On the other hand, if the buffer condition
does not satisfy the transmission requirements, namely, the buffer
entries are empty, the selected relay pair cannot help to forward
the NCS.

In this case, we drop the current relay pair $i$ and $j$ and choose
another relay pair with the second highest SINR as given by
\begin{equation}\label{ch5_equation40}
{\rm SINR^{pre}_{i,j}}={\rm SINR_{i,j}}
\end{equation}
\begin{equation}\label{ch5_equation41}
{\rm SINR_{u,v}} \in {\rm max} \{ {\rm SINR_{sr_{\Omega}}},  {\rm
SINR_{r_{\Omega}d}}\} \setminus {\rm SINR^{pre}_{i,j}},
\end{equation}
where ${\rm SINR_{u,v}}$ denotes the second highest SINR associated
with the updated relay pair $u$ and $v$. $\{ {\rm
SINR_{sr_{\Omega}}},  {\rm SINR_{r_{\Omega}d}}\} \setminus {\rm
SINR^{pre}_{i,j}}$ represents a complementary set, where we drop the
${\rm SINR^{pre}_{i,j}}$ from the set of SINR links $\{ {\rm
SINR_{sr_{\Omega}}},  {\rm SINR_{r_{\Omega}d}}\}$. Consequently, the
above process repeats until the buffer condition achieves the
transmission requirement.
\\
\\
\textbf{Reception mode}:\\ \\
If the link combination associated with
the highest relay pair SINR belongs to the source-relay phase, the
buffers are switched to reception mode. Similarly, the buffer is
checked to ensure that the corresponding buffer entries are not
full. We then have
\begin{equation}\label{ch5_equation42}
\Phi^{\rm buffer}_{i}\neq {\rm U}, \ i\in [1,2,...,L],
\end{equation}
and
\begin{equation}\label{ch5_equation43}
\Phi^{\rm buffer}_{j}\neq {\rm U}, \ j\in [1,2,...,L],
\end{equation}
where ${\rm U}$ represents a full buffer set. If the buffers are not
full, then the sources send the data to the selected relay pair $i$
and $j$ according to (\ref{ch5_equation1}). Otherwise, the algorithm
reselects a new relay pair as in (\ref{ch5_equation40}) and
(\ref{ch5_equation41}). The re-selection process stops when the
buffer entries are not full. The size $J$ of buffers plays an
important role in the system performance. When we increase the
buffer size, better channels can be selected as a relatively larger
candidate pool is generated. Therefore, extra degrees of freedom in
the system are also available.

\section{Simulations}

In this section, a simulation study of the proposed buffer-aided
PLNC scheme and linear network coding techniques is carried out. The
DS-CDMA network uses randomly generated spreading codes of length
$N=16$. The corresponding channel coefficients are modeled as
complex Gaussian random variables. We assume perfectly known
channels at the relays and receivers and remark that results with
channel estimation have the same performance hierarchy. We consider
$K=6$ users, $L=6$ relays, equal power allocation and packets with
1000 BPSK symbols in the transmissions.
For PLNC techniques, we first compare the XOR against linear network
coding schemes using different designs of the matrix $\textbf{G}$
with and without buffers ($J=4$) in Fig. \ref{ch5_fig3}. The results
show that the use of buffers can provide a significant gain in
performance
and that linear network coding techniques outperform XOR-based approaches. 

\begin{figure}[h]
\begin{center}
\def\epsfsize#1#2{0.9\columnwidth}
\epsfbox{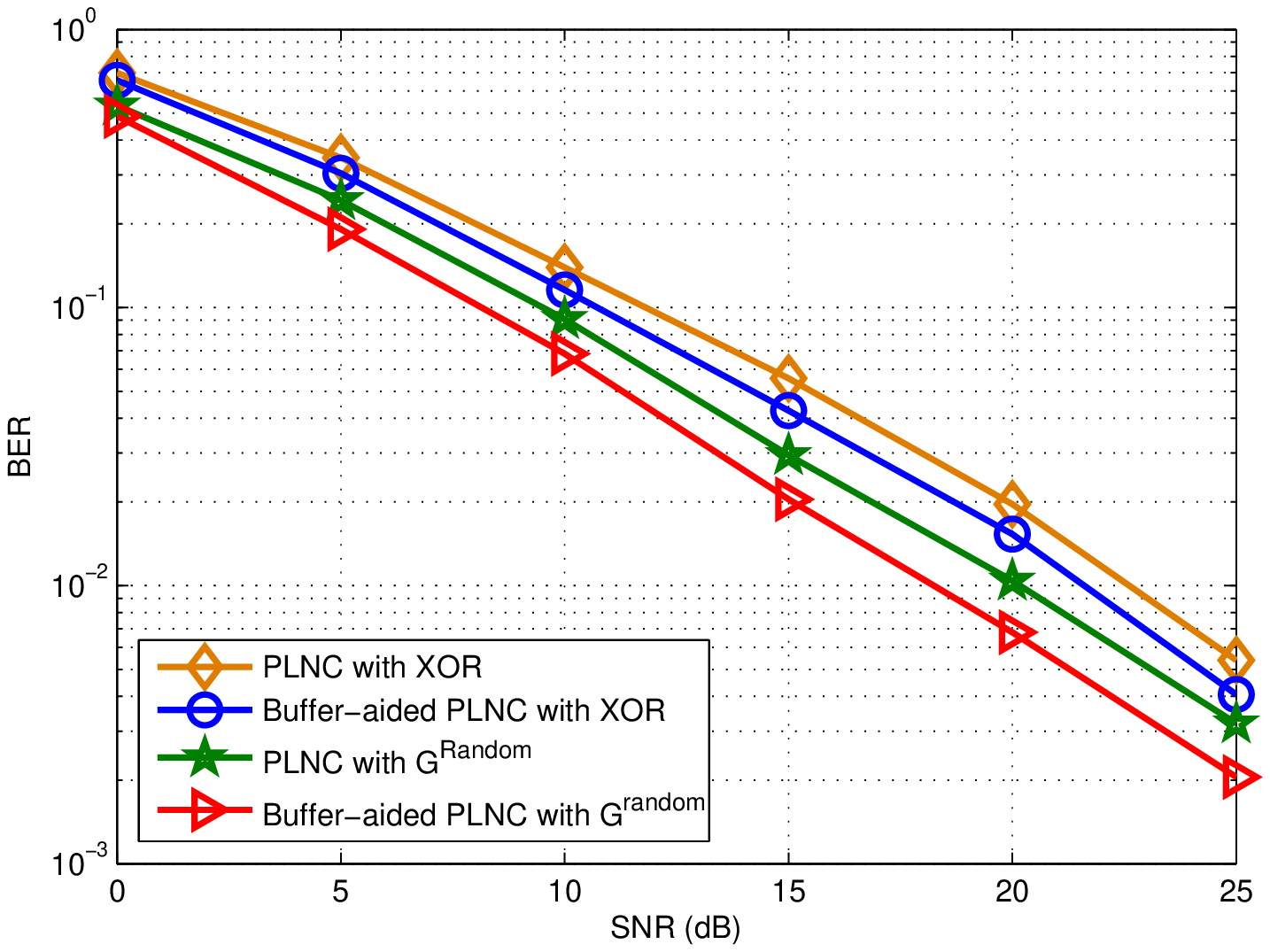} \vspace{-1.25em}\caption{Performance of linear
network coding techniques with and without buffers using BPSK
modulation.} \label{ch5_fig3}
\end{center}
\end{figure}

\begin{figure}[h]
\begin{center}
\def\epsfsize#1#2{0.9\columnwidth}
\epsfbox{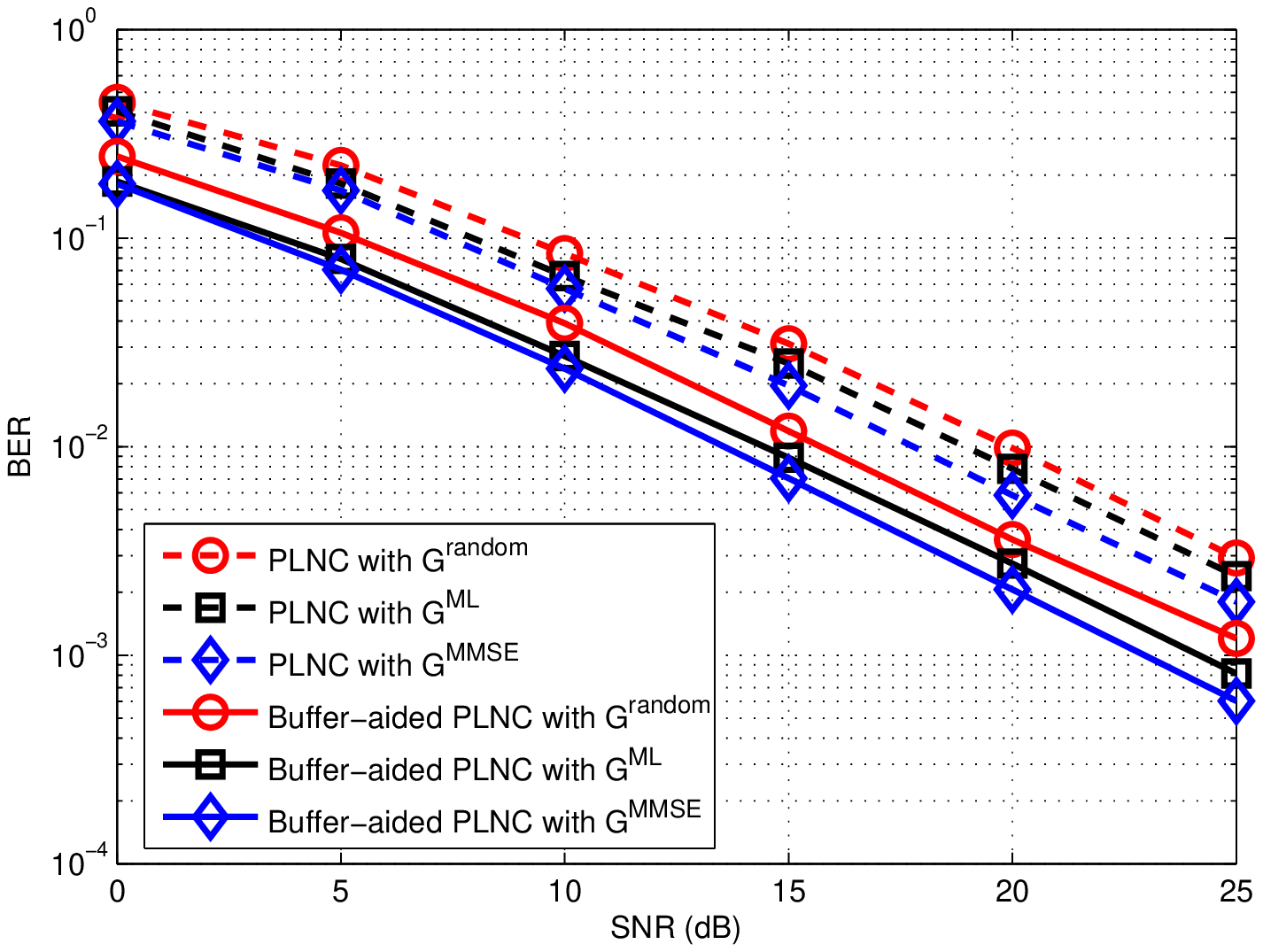} \vspace{-1.25em} \caption{Performance of linear
network coding techniques with and without buffers using BPSK
modulation.} \label{ch5_fig4}
\end{center}
\end{figure}

The BER performance of buffer-aided ($J=4$) PLNC schemes with the
proposed and existing linear network coding techniques is shown in
Fig. \ref{ch5_fig4}.  The results show that the introduction of
$\textbf{G}^{\rm MMSE}$ provides the best performance, followed by
the introduction of $\textbf{G}^{\rm ML}$ and $\textbf{G}^{\rm
random}$. In particular, the adoption of $\textbf{G}^{\rm MMSE}$
results in a gain of up to $1.5$dB in SNR over $\textbf{G}^{\rm ML}$
and up to $3$ dB in SNR as compared to $\textbf{G}^{\rm random}$ for
the same BER performance. The results also demonstrate that with the
use of buffers the overall system performance significantly
improves, achieving a gain of up to $5$ dB in terms of SNR for the
same BER performance as compared to the system without buffers even
though the diversity order is the same.

\section{Conclusions}

We have presented a buffer-aided PLNC scheme and devised novel ML
and MMSE linear network coding algorithms for cooperative DS-CDMA
systems with different relay pair selection techniques. Simulation
results show that the performance of the proposed scheme and
algorithms can offer significant gains as compared to previously
reported techniques.

{\footnotesize
\bibliographystyle{IEEEtran}
\bibliography{reference}

}
\end{document}